\documentclass[aps,floats,twocolumn,prl,nofootinbib]{revtex4-2}

\usepackage{amsfonts,amsmath,amssymb,ascmac,bm,tensor}
\usepackage{fnpct} 
\usepackage{comment}
\usepackage{ifpdf}
\usepackage{slashed}
\usepackage{color}
\usepackage[mathscr]{eucal}
\usepackage[utf8]{inputenc}
\usepackage{physics}
\usepackage{cancel}
\usepackage{soul}
\usepackage{simpler-wick}

\ifpdf
  \usepackage{graphicx}     
  \usepackage[bookmarksopen,colorlinks=true,linkcolor=bblue,citecolor=bblue,urlcolor=ppink]{hyperref}
\else     
\fi

\hypersetup{
           breaklinks=true,   
           colorlinks=true,   
           linkcolor=bblue,
           citecolor=bblue,
           urlcolor=ppink
        }

\definecolor{red}{rgb}{1,0,0}
\definecolor{darkred}{rgb}{0.6,0,0}
\definecolor{darkgreen}{rgb}{0.992447,0.623778,0.034597}
\definecolor{ppink}{rgb}{1,0.4,0.4} 
\definecolor{bblue}{rgb}{0.284602,0.317763,0.963947}
\definecolor{purple}{rgb}{0.5 ,0, 0.7}

\newcommand{\zo}{{(0)}}
\newcommand{\fo}{{(1)}}
\renewcommand{\so}{{(2)}}
\newcommand{\tho}{{(3)}}
\newcommand{\foo}{{(4)}}

\newcommand{\tmin}{\text{min}}

\newcommand{\Pl}{\text{Pl} }
\newcommand{\ee}{\text{e}}

\renewcommand{\Re}{\text{Re}}

\makeatletter
\newcommand\footnoteref[1]{\protected@xdef\@thefnmark{\ref{#1}}\@footnotemark}
\makeatother

\allowdisplaybreaks[1]

\begin{document}


\title{
Superhorizon Curvature Perturbations Are Protected against One-Loop Corrections
}

\author{Keisuke Inomata}
\affiliation{William H. Miller III Department of Physics and Astronomy, Johns Hopkins University, 3400 N. Charles Street, Baltimore, Maryland, 21218, USA}

\begin{abstract} 
\noindent
We examine one-loop corrections from small-scale curvature perturbations to the superhorizon-limit ones in single-field inflation models, which have recently caused controversy. 
We consider the case where the Universe experiences transitions of slow-roll (SR) $\to$ intermediate period $\to$ SR.
The intermediate period can be an ultra-slow-roll period or a resonant amplification period, either of which enhances small-scale curvature perturbations.
We assume that the superhorizon curvature perturbations are conserved at least during each of the SR periods.
Within this framework, we show that the superhorizon curvature perturbations during the first and the second SR periods coincide at one-loop level in the slow-roll limit.
\end{abstract}

\date{\today}
\maketitle

\noindent
\emph{{\bf Introduction.}}
Over the past few decades, the cosmological perturbations have been measured through the large scale structure (LSS) and the anisotropies of the cosmic microwave background (CMB) and have given us a lot of insights into the early Universe. 
The curvature perturbations are often used to characterize the amplitudes of the cosmological perturbations. 
This is mainly because the curvature perturbations are considered to become constant (that is, conserved) once they exit the horizon during the inflation in the absence of isocurvature perturbations~\cite{Bardeen:1980kt,Kodama:1984ziu}.
In particular, superhorizon curvature perturbations had been thought to be conserved in single-field inflation models, which do not predict isocurvature perturbations.
In this work, we focus on the conservation of the superhorizon curvature perturbations in single-field inflation models and simply call it the conservation of the curvature perturbations hereafter.

Intriguingly, Ref.~\cite{Kristiano:2022maq} has recently claimed the non-conservation of the curvature perturbations at one-loop order in the in-in (Schwinger–Keldysh) formalism, which has often been used to examine the higher(nonlinear)-order corrections to the power spectrum~\cite{Weinberg:2005vy,Senatore:2009cf}. 
Specifically, it claims that the ultra-slow-roll (USR) period, sandwiched between slow-roll (SR) periods, during inflation results in the non-conservation of the curvature perturbations.
\footnote{See Ref.~\cite{Cheng:2021lif} for an earlier work that claims the non-conservation of the curvature perturbations with the use of the Hartree factorization.
See also Refs.~\cite{Inomata:2022yte,Fumagalli:2023loc} for the loop corrections to the power spectrum around the scales where the linear perturbations are enhanced.
}
This setup (SR $\to$ USR $\to$ SR) is a typical single-field scenario for primordial black hole (PBH) production.
The possibility of the non-conservation indicates that large-(such as CMB/LSS-)scale perturbations, which exit the horizon earlier during inflation, are affected by the enhanced small-scale perturbations that exit the horizon later during inflation and produce PBHs after inflation~\cite{Zeldovich:1967lct,Hawking:1971ei,Carr:1974nx,Carr:1975qj}.
If the one-loop power spectrum is larger than the tree-level one on the CMB/LSS scales, the perturbation theory is no longer reliable, which puts strong constraints on inflation models for PBH production~\cite{Kristiano:2022maq}.
We remark that PBHs have attracted attention also from particle physics and astrophysics because they are candidates for dark matter~\cite{Chapline:1975ojl,Ivanov:1994pa,Yokoyama:1995ex,GarciaBellido:1996qt,Afshordi:2003zb,Frampton:2010sw,Belotsky:2014kca,Carr:2016drx,Inomata:2017okj,Espinosa:2017sgp} and/or for the BHs detected by LIGO-Virgo-KAGRA collaborations~\cite{Bird:2016dcv,Clesse:2016vqa,Sasaki:2016jop,Eroshenko:2016hmn,Sasaki:2018dmp,Green:2020jor,Escriva:2022duf}.

However, the claim of Ref.~\cite{Kristiano:2022maq} seems inconsistent with the separate universe assumption: a sufficiently large-scale region, which is at least larger than the horizon scale, can be regarded as a (locally) homogeneous and isotropic universe and evolves independently of other regions~\cite{Sasaki:1998ug,Wands:2000dp,Lyth:2003im}.
With this assumption, Ref.~\cite{Lyth:2004gb} has shown the curvature conservation at the non-perturbative level.
In this sense, Ref.~\cite{Kristiano:2022maq} indirectly claims the possibility that the separate universe assumption becomes invalid due to the USR period.
Note that, although it is well-known that the curvature perturbations that are on superhorizon but close to the horizon grow during a USR period~\cite{Kinney:1997ne,Inoue:2001zt,Kinney:2005vj,Martin:2012pe}, the non-conservation claimed in Ref.~\cite{Kristiano:2022maq} also applies to the curvature perturbations in the superhorizon limit, where the linear curvature perturbations are conserved.
We here stress that the violation of the separate universe assumption infers the universes connected over a distance larger than the horizon scales, which violates the causality.
In this sense, investigating the claim of the non-conservation of the curvature is vital not only in Cosmology, but also in the basic of physics.

The result of Ref.~\cite{Kristiano:2022maq} has been followed up by many papers~\cite{Riotto:2023hoz,Choudhury:2023vuj,Choudhury:2023jlt,Kristiano:2023scm,Riotto:2023gpm,Firouzjahi:2023aum,Motohashi:2023syh,Firouzjahi:2023ahg,Franciolini:2023agm,Tasinato:2023ukp,Cheng:2023ikq,Fumagalli:2023hpa,Maity:2023qzw,Tada:2023rgp,Firouzjahi:2023bkt,Davies:2023hhn,Iacconi:2023ggt,Saburov:2024und}.
Most of the follow-up works are consistent with the statement that, if the transition from USR to SR is sudden compared to the Hubble timescale at that time, the one-loop power spectrum becomes $\mathcal P_{\zeta, 1\text{-loop}}(q) \gtrsim \mathcal O(\mathcal P_\zeta(q) \mathcal P_{\zeta}(k_\text{peak}))$ after the USR period, where $q \ll k_\text{peak}$ with $k_\text{peak}$ being the scale for the global maximum of the enhanced power spectrum, close to the horizon scale at the beginning of the USR period.
In short, many papers claim that the non-conservation of the superhorizon curvature perturbations is possible.

On the other hand, two papers disagree with the non-conservation of the curvature perturbations.
The first paper is Ref.~\cite{Fumagalli:2023hpa}, where the author takes into account the boundary terms that are not in the analysis of Ref.~\cite{Kristiano:2022maq} and found that the one-loop corrections disappear in the superhorizon limit.
However, Ref.~\cite{Firouzjahi:2023bkt} pointed out that the analysis in the paper misses the nonzero commutation relation between the curvature perturbations at different times.
The second paper is Ref.~\cite{Tada:2023rgp}, which claims that the cancellation of the one-loop power spectrum is guaranteed from Maldacena's consistency relation~\cite{Maldacena:2002vr}. 
However, the renormalization scheme taken in Ref.~\cite{Tada:2023rgp} is questioned in Ref.~\cite{Firouzjahi:2023bkt}. 
We also note that these two papers~\cite{Fumagalli:2023hpa,Tada:2023rgp} only focus on the one-loop contribution from the cubic interaction of the perturbations, while the quartic interaction can give a comparable one-loop contribution~\cite{Firouzjahi:2023aum,Maity:2023qzw}.

The important feature of the situation in Ref.~\cite{Kristiano:2022maq} and the subsequent works is the existence of a USR period, sandwiched by SR periods. 
This motivates us to examine the one-loop corrections in the Universe that experiences transitions of SR $\to$ intermediate period $\to$ SR in this work. 
We assume that, during the intermediate period, the amplitudes of small-scale curvature perturbations are enhanced. 
This is a typical PBH scenario within single-field inflation models.
For instance, the intermediate period can be a USR period or a period for the resonant amplification of the perturbations due to oscillatory features in the inflaton potential~\cite{Inomata:2022yte}.
We note that our setup covers the case with a sudden transition from USR to SR.
We additionally assume that the separate universe assumption is valid at least during the SR periods, which secures the conservation of the curvature perturbations at non-perturbative (including one-loop) level at least during each of those periods~\cite{Lyth:2004gb}.
Within this setup, we will show that the superhorizon curvature perturbations during the first and the second SR periods coincide at the one-loop level in the slow-roll limit.

Before moving to equations, we highlight the key differences from the previous works, especially from those claiming the curvature conservation~\cite{Fumagalli:2023hpa,Tada:2023rgp}.
As discussed in Ref.~\cite{Maldacena:2002vr}, the higher-order action, needed for one-loop calculation, is often derived in two gauges. 
One is comoving gauge, taken in the previous works, and the other is spatially-flat gauge, taken in this work.
Although the final results should not depend on the gauge, the calculation complexity does.
In particular, the higher-order action in comoving gauge includes boundary terms, whose roles in one-loop calculation are under debate~\cite{Fumagalli:2023hpa,Tada:2023rgp,Firouzjahi:2023bkt,Braglia:2024zsl,Kawaguchi:2024lsw}.
On the other hand, as we will see below, the higher-order action in spatially-flat gauge does not have boundary terms in the slow-roll limit, which simplifies the calculation.
This simplification also enables us to easily take into account the quartic interaction, which is ignored in the previous works~\cite{Fumagalli:2023hpa,Tada:2023rgp}.

Throughout this work, we discuss the gauge independent quantity $\zeta$ that coincides the curvature perturbation in uniform-density gauge.
Note that, in the superhorizon limit, $\zeta$ coincides even at the nonperturbative level with another gauge independent quantity $\mathcal R$ that coincides the curvature in comoving gauge~\cite{Lyth:2004gb}. 
Also, we neglect the contributions from tensor and vector perturbations to focus on the loop corrections from the enhanced curvature perturbations on small scales.

\vspace{5pt}
\noindent
\emph{{\bf Key equations of motion.}}
We work with the inflaton fluctuation $\delta \phi$ in spatially-flat gauge~\cite{Maldacena:2002vr,Pajer:2016ieg}, which allows us to neglect the metric perturbations in the slow-roll limit, $\epsilon(\equiv - \dot H/H^2) \to 0$ with some quantities fixed.
We will explain the slow-roll limit in detail below Eq.~(\ref{eq:f_eom}).
Within that gauge, we consider a canonical single field, whose action and Lagrangian density are given by 
\begin{align}
  S = \int \dd \eta \, \dd^3 x \, a^4 \mathcal L, \ \  \mathcal L = - \frac{1}{2} \partial^\mu \phi \partial_\mu \phi - V(\phi).
\end{align}
From this, we can obtain the equation of motion:
\begin{align}
  \phi'' + 2 \mathcal H \phi' - \nabla^2 \phi  + a^2 V_\fo(\phi) = 0,
  \label{eq:eom}
\end{align}
where the prime denotes the conformal time derivative, $\mathcal H \equiv a'/a$, and $V_{(n)} \equiv \partial^n V(\phi)/\partial \phi^n$.

We here express the scalar field as 
\begin{align}
  \phi = \bar \phi + \delta \phi^\fo + \delta \phi^\so + \delta \phi^\tho + \cdots,
\end{align}
where $\bar \phi$ is the background value, the superscript for $\delta \phi$ denotes the order of the perturbations, and $\cdots$ denotes the higher-order perturbations.
From Eq.~(\ref{eq:eom}), the equation of motion for the background up to one loop level becomes 
\begin{align}
  \bar \phi'' + 2 \mathcal H \bar \phi' + a^2 V_\fo(\bar \phi) = - \frac{a^2}{2}V_\tho(\bar \phi) \expval{(\delta \phi^{\fo})^2}, 
  \label{eq:back_eom}
\end{align}
where the bracket means the ensemble average, which leads to $\expval{\phi} = \bar \phi$, and the right-hand side (RHS) is the one-loop backreaction from the perturbations.
For convenience, we here rewrite this equation of motion with the physical time:
\begin{align}
  \ddot {\bar \phi} + 3 H \dot{\bar \phi} + V_\fo(\bar \phi) = - \frac{1}{2}V_\tho(\bar \phi) \expval{(\delta \phi^{\fo})^2}, 
  \label{eq:back_eom_phys}
\end{align}
where the dot denotes the physical time derivative.
We here take the physical time derivative of this equation and partially reexpress it with the conformal time:
\begin{align}
  &\hat {\mathcal N}_0\bar \Pi
  = - \frac{a^2}{2} \left( V_\tho \expval{\delta \phi^2}^{\bm{\cdot}} + V_\foo \expval{\delta \phi^2} \bar \Pi \right), 
  \label{eq:pi_eom}
\end{align}
where $\bar \Pi \equiv \dot{\bar \phi}$, 
\begin{align}
  \hat{\mathcal N}_k \equiv \frac{\partial^2}{\partial \eta^2} + 2 \mathcal H \frac{\partial}{\partial \eta} + k^2 + a^2 V_\so(\bar \phi),
  \label{eq:n_0_def}
\end{align}
and we have neglected the slow-roll suppressed term proportional to $a^2 \dot H \, \bar \Pi$. 
Note that we use an equal sign ``$=$'' with slow-roll suppressed terms neglected throughout this work.
We will see $\hat {\mathcal N}_{k \neq 0}$ just below.
The RHS of Eq.~(\ref{eq:pi_eom}) is the one-loop backreaction from the perturbations to the evolution of $\bar \Pi$.
For convenience, we expand $\bar \Pi$ with respect to the perturbation order as $\bar \Pi = \bar \Pi^\zo + \bar \Pi^\so + \cdots$, which satisfy
\begin{align}
  \label{eq:eom_pi_zo}
    &\hat{\mathcal N}_0 \bar \Pi^\zo = 0, \\
    &\hat {\mathcal N}_0\bar \Pi^\so
   = - \frac{a^2}{2} \left(V_\tho \expval{(\delta \phi^\fo)^2}^{\bm{\cdot}} + V_\foo \expval{(\delta \phi^\fo)^2} \bar \Pi^\zo \right).
  \label{eq:eom_pi_so}
\end{align}

Expanding Eq.~(\ref{eq:eom}) with respect to $\delta \phi$, we obtain the equation of motion for the perturbations in the Fourier space:
\begin{align}
    \label{eq:eom_fo}
    &\hat {\mathcal N}_k \delta \phi^\fo_{\bm k}= 0, \\
    \label{eq:eom_so}    
    &\hat {\mathcal N}_k \delta \phi^\so_{\bm k}= -\frac{a^2}{2}V_\tho \int \frac{\dd^3 p}{(2\pi)^3} \delta \phi^{\fo}_{\bm k - \bm p} \delta \phi^{\fo}_{\bm p}, \\ 
    \label{eq:eom_tho}
    &\hat {\mathcal N}_k \delta \phi^\tho_{\bm k}= -\frac{a^2}{2}V_\tho \int \frac{\dd^3 p}{(2\pi)^3} (\delta \phi^{\fo}_{\bm k - \bm p} \delta \phi^\so_{\bm p} + \delta \phi^{\so}_{\bm k- \bm p} \delta \phi^\fo_{\bm p}) \nonumber \\
    &\quad
    - \frac{a^2}{6} V_\foo \int \frac{\dd^3 p}{(2\pi)^3} \int \frac{\dd^3 p'}{(2\pi)^3} \delta \phi^\fo_{\bm p} \delta \phi^\fo_{\bm p'} \delta \phi^\fo_{\bm k - \bm p- \bm p'}.
\end{align}
Comparing Eqs.~(\ref{eq:eom_pi_zo}) and (\ref{eq:eom_fo}), we can see that the tree-level background $\bar \Pi^\zo$ evolves in the same way as the superhorizon-limit $\delta \phi^\fo$ ($\lim_{k \to 0} \delta \phi_{\bm k}$) does, where we have set their decaying modes during the first SR period to zero.
From this, we can also see that the linear superhorizon-limit curvature perturbation, $\zeta^\fo = -\delta \phi^\fo/(H^{-1}\bar \Pi^\zo)$, is constant. 
The word ``superhorizon'' hereafter means the scales much larger than the horizon scale on which the first-order curvature perturbations are conserved even if the intermediate period is a USR.

For later convenience, we also summarize the equations for $\delta \Pi \equiv \delta \dot \phi$ and $f_{\bm k} \equiv \partial \delta \phi_{\bm k}/\partial (\ln k)$.
By taking the derivative of Eq.~(\ref{eq:eom_fo}) with respect to time and wavenumber and neglecting slow-roll suppressed terms, we obtain the following equations:
\begin{align}
  \hat {\mathcal N}_k \delta \Pi^\fo_{\bm k} &= 2k^2 H \delta \phi^\fo_{\bm k} - a^2 V_\tho \bar \Pi^\zo \delta \phi^\fo_{\bm k},
  \label{eq:dpi_eom}\\
    \hat {\mathcal N}_k f^\fo_{\bm k} &= -2k^2 \delta \phi^\fo_{\bm k}.
  \label{eq:f_eom}
\end{align}

Before moving to the one-loop calculation, let us explain the slow-roll limit that we take in this work. 
We consider the limit of $\epsilon \to 0$ with two quantities fixed. 
The first fixed quantity is $H^2/(\epsilon M_\Pl^2)$. We fix it to keep the amplitude of the curvature perturbations. 
The second fixed quantity is the timescale of the change of the potential derivative terms during the intermediate period.
More specifically, we consider that the features in the potential derivatives, which induce the loop contributions, are related to $\epsilon$ as $V_{(n)} \propto \mathcal O(\epsilon^{(2-n)/2})$.
For example, to realize the transition from USR to SR through $\Delta N$ e-folds when $\epsilon \simeq \epsilon_0$, the potential derivatives should be $V_{(n)}/(H^2 M_\Pl^2) \simeq \mathcal O(\epsilon_0^{(2-n)/2}/(\Delta N^{n-1} M_\Pl^{n}))$ during the transition.
In this case, we fix $\Delta N$ during the limit of $\epsilon \to 0$.
See also Ref.~\cite{Inomata:2022yte} for similar relations in the case of the oscillatory features in the potential.
This limit corresponds to the conformal (or decoupling) limit in the effective field theory of inflation~\cite{Pajer:2016ieg,Baumann:2011su}.
In this limit, we can safely neglect the metric perturbations, compared to the potential derivative terms~\cite{Pajer:2016ieg,Inomata:2022yte}.

\vspace{5pt}
\noindent
\emph{{\bf One loop.}}
One-loop power spectrum has been calculated in spatially-flat gauge with the in-in formalism in Ref.~\cite{Inomata:2022yte}, which shows that the one-loop power spectrum can be written as 
\begin{align}
  \expval{\delta \phi_{\bm q} \delta \phi_{\bm q'}}_{1\text{-loop}} &= \expval{\delta \phi^\so_{\bm q} \delta \phi^\so_{\bm q'}} \nonumber \\ 
  &\quad + \expval{\delta \phi^\fo_{\bm q} \delta \phi^\tho_{\bm q'}} + \expval{\delta \phi^\tho_{\bm q} \delta \phi^\fo_{\bm q'}}.
  \label{eq:1_loop}
\end{align}
We can calculate $\delta \phi$ by solving Eqs.~(\ref{eq:eom_fo})-(\ref{eq:eom_tho}).
We stress that the RHS in Eq.~(\ref{eq:1_loop}) is obtained with the in-in formalism (see Ref.~\cite{Musso:2006pt} and Appendix in Ref.~\cite{Inomata:2022yte}). 
In the following, we focus on the one-loop power spectrum in the superhorizon limit and use the character of $\bm q$ (or $\bm q'$) for the superhorizon-limit modes.
In addition, we consider the case where the one-loop power spectrum is dominantly induced by the perturbations whose scales are much smaller than that for the one-loop power spectrum.

In our setup, the dominant contribution of $\expval{\delta \phi^\so_{\bm q} \delta \phi^\so_{\bm q'}}$ is independent of $\delta \phi^\fo_{\bm q}$ because $\delta \phi^\so_{\bm q}$ is dominantly induced by the square of the linear perturbations on the small scales.
Then, the causality makes this contribution Poisson-like and results in the volume-suppression ($\mathcal P_{\delta \phi}(q) \propto q^3$) in the superhorizon limit~\cite{Riotto:2023hoz}.
Given this, we neglect this contribution in the following.
On the other hand, we must be careful about the other contributions, $\expval{\delta \phi^\fo_{\bm q} \delta \phi^\tho_{\bm q'}} + \expval{\delta \phi^\tho_{\bm q} \delta \phi^\fo_{\bm q'}}$, because $\delta \phi^\tho_{\bm q}$ depends on $\delta \phi^\fo_{\bm q}$, as we will see below.
In the following, we focus on the evolution of $\delta \phi^\tho_{\bm q}$ because its behavior determines whether the one-loop power spectrum remains in the superhorizon limit. 
For convenience, we reexpress Eq.~(\ref{eq:eom_tho}) as 
\begin{align}
  \hat {\mathcal N}_0 \delta \phi^\tho_{\bm q} = \hat {\mathcal N}_0 \delta \phi^\tho_{\bm q}|_{V_\tho} + \hat {\mathcal N}_0 \delta \phi^\tho_{\bm q}|_{V_\foo},
  \label{eq:phi_tho}
\end{align}
where the first and the second terms in the RHS correspond to the first and the second lines in the RHS of Eq.~(\ref{eq:eom_tho}), respectively.
We hereafter assume that $V_\tho$ and $V_\foo$ are zero during the SR periods for simplicity.

Let us begin with the second term in Eq.~(\ref{eq:phi_tho}) because it is simpler. 
It can be expressed as  
\begin{align}
  \hat {\mathcal N}_0 \delta \phi^\tho_{\bm q}|_{V_\foo} &= -\frac{a^2}{2} V_\foo \expval{(\delta \phi^{\fo})^2} \delta \phi^\fo_{\bm q},
  \label{eq:phi_tho_foo}
\end{align}
where $\expval{(\delta \phi^\fo)^2} \simeq \int^{\Lambda_* \frac{a(\eta)}{a(\eta_*)}}_{k_\tmin} \dd \ln k\, \mathcal P_{\delta \phi^\fo}(k)$ with $k_\tmin (\gg q)$ being the large-scale (IR) cutoff wavenumber.\footnote{
We can also set a physical IR cutoff in the wavenumber integral of $\expval{(\delta \phi^\fo)^2}$, instead of a comoving one. In that case, the physical IR cutoff scale must be larger than the physical scale for $k_\tmin$ (smaller in $k$) at the end of the intermediate period. }
We assume that the perturbations on $k > k_\tmin$ dominantly contribute to the one loop and those on $k < k_\tmin$ are negligible.
The $\Lambda_* \frac{a(\eta)}{a(\eta_*)}$ is a physical UV cutoff scale with $\Lambda_*$ being the UV cutoff scale at some fiducial time $\eta_*$.

Next, we discuss the first term in Eq.~(\ref{eq:phi_tho}). 
To this end, we first calculate $\delta \phi^\so$ by solving Eq.~(\ref{eq:eom_so}) with the Green function method. 
The Green function for Eq.~(\ref{eq:eom_so}) is defined as the function satisfying $\hat {\mathcal N}_k g_k (\eta;\eta') = \delta(\eta- \eta')$.
Using this, we obtain 
\begin{align}
  \delta \phi^\so_{\bm k}(\eta) =& -\int^\eta_{\eta_i} \dd \eta' g_k(\eta;\eta') \frac{a^2}{2} V_\tho \int \frac{\dd^3 p}{(2\pi)^3} \delta \phi^{\fo}_{\bm k - \bm p} \delta \phi^{\fo}_{\bm p},
  \label{eq:d_phi_so}
\end{align}
where $\eta_i$ is some time before the intermediate period and the terms without time arguments inside the time integral depend on $\eta'$, not on $\eta$.
As the initial condition, we have taken $\delta \phi^{(n>1)} = 0$ during the first SR period by assuming $\delta \phi$ is a Gaussian variable then.
For concreteness, we relate $\delta \phi_{\bm k}$ to the operators as 
\begin{align}
  &\delta \phi^\fo(\bm x,\eta) = \int \frac{\dd^3 k}{(2\pi)^3} \ee^{i \bm k \cdot \bm x} \delta \phi^\fo_{\bm k}(\eta) \nonumber \\
  &= \int \frac{\dd^3 k}{(2\pi)^3} \ee^{i \bm k \cdot \bm x} \left[ U_k(\eta)\hat a(\bm k) +  U^{*}_k(\eta) \hat a^{\dagger}(-\bm k) \right],
\end{align}
where the commutation relations between the creation and annihilation operators are given by $[\hat a(\bm k), \hat a(\bm k')] = [\hat a^\dagger(\bm k), \hat a^\dagger(\bm k')] = 0$ and $[\hat a(\bm k), \hat a^\dagger(-\bm k')] = (2\pi)^3 \delta(\bm k + \bm k')$.
Substituting Eq.~(\ref{eq:d_phi_so}) into Eq.~(\ref{eq:eom_tho}) and using this expression of $\delta \phi_{\bm k}$, we obtain 
\begin{align}
  \hat {\mathcal N}_0 \delta \phi^\tho_{\bm q}|_{V_\tho} &= a^2 V_\tho \int \frac{\dd^3 k}{(2\pi)^3}\int^\eta_{\eta_i} \dd \eta' g_k(\eta;\eta') a^2 V_\tho \nonumber \\
  & \qquad \times
  \Re[U_k(\eta)U^*_k(\eta')] \delta \phi^\fo_{\bm q},
  \label{eq:phi_tho_tho}  
\end{align}
where we have taken the ensemble average for the square of the source perturbations and extracted $\delta \phi^\fo_{\bm q}$ because we are interested in $\expval{\delta \phi^\fo_{\bm q} \delta \phi^\tho_{\bm q'}} + \expval{\delta \phi^\tho_{\bm q} \delta \phi^\fo_{\bm q'}}$.
We have also used $g_{|\bm k - \bm q|}(\eta;\eta') \to g_{k}(\eta;\eta')$ in $q/k \to 0$ to obtain this expression.

In the following, we derive the key equation, $\delta \phi^\tho_{\bm q} = (\bar \Pi^\so/\bar \Pi^\zo)\delta \phi^\fo_{\bm q}$.
To calculate the first term in Eq.~(\ref{eq:eom_pi_so}), let us get the expression of $\delta \Pi (\equiv \delta \dot \phi)$ first.
From Eqs.~(\ref{eq:dpi_eom}) and (\ref{eq:f_eom}), we obtain 
\begin{align}
\label{eq:pi_fo}
  \delta \Pi^\fo_{\bm k}(\eta) &=  -\int^\eta_{\eta_i} \dd \eta' g_k(\eta;\eta') \left[a^2 V_\tho \bar \Pi^\zo \delta \phi^\fo_{\bm k} \right] - H f^\fo_{\bm k} \nonumber \\
  &\qquad
  + C_{\bm k} H U_k(\eta) + D_{\bm k} H U^*_k(\eta),
\end{align}
where $C_{\bm k}$ and $D_{\bm k}$ are some constants, which include the creation and the annihilation operators. 
Let us determine these constants by taking into account the initial condition.
$\delta \Pi^\fo_{\bm k}$ and $f_{\bm k}$ can be expressed as 
\begin{align}
  \label{eq:pi_ini}
  \delta \Pi_{\bm k}^\fo &= \frac{U_k'(\eta)\hat a(\bm k) +  U^{*\prime}_k(\eta) \hat a^{\dagger}(-\bm k)}{a(\eta)}, \\
  f^\fo_{\bm k} &= \frac{\partial}{\partial \ln k}\left[U_k(\eta)\hat a(\bm k) +  U^{*}_k(\eta) \hat a^{\dagger}(-\bm k)\right].
  \label{eq:f_k}
\end{align}
To proceed, we here assume that all modes are in the Bunch-Davies vacuum solution at least for $\eta \leq \eta_i$, which leads to~\cite{Baumann:2009ds}
\begin{align}
  U_k(\eta \leq \eta_i) = i\frac{H}{\sqrt{2 k^3}}(1+ i k\eta) \ee^{-i k\eta}.
\end{align}
From this, we can see that $k^{3/2} U_k(\eta)$ is the function only of $k\eta$ in $\eta \leq \eta_i$ and the limit of $\epsilon \to 0$.
Then, we can obtain the following relation:
\begin{align}
  \frac{U'_k(\eta \leq \eta_i)}{a(\eta)} = - \frac{H}{k^{3/2}} \frac{\partial(k^{3/2} U_k(\eta \leq \eta_i))}{\partial\ln k},
\end{align}
and the same for $U^*_k$.
Substituting this into Eq.~(\ref{eq:pi_ini}) and comparing Eqs.~(\ref{eq:pi_fo}) and (\ref{eq:f_k}), we finally obtain $C_{\bm k} = -\frac{3}{2}\hat a(\bm k) + \frac{\partial \hat a(\bm k)}{\partial \ln k}$ and $D_{\bm k} = -\frac{3}{2}\hat a^\dagger(-\bm k) + \frac{\partial \hat a^\dagger(-\bm k)}{\partial \ln k}$.
Then, we obtain 
\begin{align}
&\expval{(\delta \phi^\fo(\eta))^2}^{\bm{\cdot}} \nonumber \\ 
&= -2 \int \frac{\dd^3 k}{(2\pi)^3}\int^\eta_{\eta_i} \dd \eta' g_k(\eta;\eta') a^2 V_\tho \bar\Pi^\zo \Re[U_k(\eta)U^*_k(\eta')] \nonumber \\
& \  - H \int^{\Lambda_* \frac{a(\eta)}{a(\eta_*)}}_{k_\tmin} \frac{\dd k}{k}\, \frac{\dd \mathcal P_{\delta \phi^\fo}(k,\eta)}{\dd \ln k} + H \mathcal P_{\delta \phi^\fo}\left(\Lambda_* \frac{a(\eta)}{a(\eta_*)},\eta\right),
\label{eq:dphi2_dot}
\end{align}
where $\mathcal P_{\delta \phi^\fo}(k,\eta) = k^3|U_k(\eta)|^2/(2\pi^2)$. 
The contributions except for the last term in the second line can be calculated from $\expval{\delta \Pi^\fo(\eta) \delta \phi^\fo(\eta)} + \expval{\delta \phi^\fo(\eta) \delta \Pi^\fo(\eta)}$ with the UV cutoff fixed.
The last term in the second line comes from the time derivative of the physical UV cutoff in $\expval{(\delta \phi^\fo)^2} = \int^{\Lambda_* \frac{a(\eta)}{a(\eta_*)}} \dd \ln k\, \mathcal P_{\delta \phi^\fo}(k)$.
The second line eventually becomes $H \mathcal P_{\delta \phi}(k_\tmin)$ and this contribution is negligible by definition of $k_\tmin$. 
Using these, we can reexpress Eq.~(\ref{eq:eom_pi_so}) as 
\begin{align}
  \hat {\mathcal N}_0\bar \Pi^\so
  &= a^2 V_\tho \int \frac{\dd^3 k}{(2\pi)^3}\int^\eta_{\eta_i} \dd \eta' g_k(\eta;\eta') a^2 V_\tho \nonumber \\
  & \qquad \qquad \qquad \qquad \qquad 
  \times \Re[U_k(\eta)U^*_k(\eta')] \bar \Pi^\zo \nonumber \\
  & \quad -\frac{a^2}{2} V_\foo \expval{(\delta \phi^{\fo})^2} \bar \Pi^\zo.
\end{align}
Comparing this expression with the equation of motion for $\delta \phi^\tho_{\bm q}$ (Eqs.~(\ref{eq:phi_tho}), (\ref{eq:phi_tho_foo}), and (\ref{eq:phi_tho_tho})), we can see that this equation for $\bar \Pi^\so$ is the same as that for $\delta \phi^\tho_{\bm q}$ in the exchange of $\bar \Pi^\zo \leftrightarrow \delta \phi^\fo_{\bm q}$.
Recalling that $\delta \phi^\fo_{\bm q}/\bar \Pi^\zo$ is always constant, we finally obtain the key equation, 
\begin{align}
  \delta \phi^\tho_{\bm q} = \frac{\bar \Pi^\so}{\bar \Pi^\zo} \delta \phi^\fo_{\bm q},
  \label{eq:d_tho_d_fo}
\end{align}
where we have set $\bar \Pi^\so = \delta \phi^\tho_{\bm q} = 0$ during the first SR period as the initial condition.
From this, the following relation is satisfied during all the periods (the SR, the intermediate, and the transition periods between them): 
\begin{align}
  \left.\frac{\delta \phi_{\bm q}}{H^{-1}\bar \Pi}\right|_{\leq 1\text{-loop}} \equiv \frac{\delta \phi^\fo_{\bm q} + \delta \phi^\so_{\bm q} + \delta \phi^\tho_{\bm q}}{H^{-1}(\bar\Pi^\zo + \bar\Pi^\so)} = \frac{\delta \phi^\fo_{\bm q}}{H^{-1}\bar \Pi^\zo}= -\zeta^\fo_{\bm q}, 
  \label{eq:one_loop_cons}
\end{align}
where, in the first equality, we have used Eq.~(\ref{eq:d_tho_d_fo}) and dropped $\delta \phi^\so$ because of the reason explained between Eqs.~(\ref{eq:1_loop}) and (\ref{eq:phi_tho}). 
Note again that $\zeta^\fo$ is constant in the superhorizon limit.
Eq.~(\ref{eq:one_loop_cons}) is the main result of this work. 
We stress that Eq.~(\ref{eq:one_loop_cons}) does not depend on details of the intermediate period itself and the transition between the intermediate period and the SR periods.

Let us check how the left-hand side of Eq.~(\ref{eq:one_loop_cons}) is related to the curvature perturbation at one-loop level.
To connect $\delta \phi$ and $\zeta$, we finally use our assumption that the separate universe assumption is valid and the curvature perturbations are conserved at the nonperturbative (including one-loop) level during the SR periods~\cite{Lyth:2004gb}. 
Let us tentatively further assume that the inflaton potential for the first and the second SR periods has linear-potential regions with $V \simeq V_0(1 + \sqrt{2\epsilon_i}\phi/M_\Pl)$, respectively, where $i \in \{1,2\}$ and $\epsilon_1$ and $\epsilon_2$ are the slow-roll parameters when the inflaton slowly rolls on the regions during the first and the second SR periods, respectively.
When the inflaton slowly rolls on those regions, we can simply connect the curvature perturbation to the inflaton fluctuation at the one-loop level by using the $\delta N$ formalism, $\zeta = \delta N = -\delta \phi/(H^{-1}\bar \Pi)$. 
Then, from Eq.~(\ref{eq:one_loop_cons}), we can conclude that the superhorizon curvature perturbations during the first and the second SR periods coincide at the one-loop level.
This conclusion also applies to the inflaton potential without the linear-potential regions because our assumption about the conservation of the curvature perturbations secures that the curvature perturbations during the SR periods are independent of the existence of the linear-potential regions.

We finally mention the renormalization of the loop contributions. 
In general, we need to add the counter terms that cancel the divergences in loop contributions. 
In our calculation, we can add the counter term as $V_{\so} \rightarrow V_{\so} + V_{\so, c}$ with $V_{\so, c}$ canceling the divergences. 
The point is that the counter term is added automatically in the same way in both the equations of motion for $\bar \Pi^\so$ and $\delta \phi^\tho_{\bm q}$, which keeps Eqs.~(\ref{eq:d_tho_d_fo}) and (\ref{eq:one_loop_cons}).

\vspace{5pt}
\noindent
\emph{{\bf Conclusion.}}
We have calculated the one-loop corrections from enhanced small-scale perturbations to the superhorizon-limit curvature perturbations in the single-field inflation models where the Universe experiences transitions of SR $\to$ intermediate period $\to$ SR. 
The enhancement of the small-scale perturbations occurs during the intermediate period, which is a typical PBH scenario within single-field inflation models.
The intermediate period can be a USR period or a period of the resonant amplification of the perturbations.
To proceed, we have assumed that the separate universe assumption is valid and the superhorizon curvature perturbations are conserved at the nonperturbative level at least during each of the SR periods.
Then, we found that the superhorizon-limit curvature perturbations during the first and the second SR periods coincide at the one-loop level in the slow-roll limit.
This shows the conservation of the superhorizon curvature perturbations at the one-loop level.
We stress that our result applies to the case with a sudden transition from USR to SR, for which the non-conservation of the superhorizon curvature perturbations was claimed in the previous papers~\cite{Kristiano:2022maq,Riotto:2023hoz,Choudhury:2023vuj,Choudhury:2023jlt,Kristiano:2023scm,Riotto:2023gpm,Firouzjahi:2023aum,Motohashi:2023syh,Firouzjahi:2023ahg,Franciolini:2023agm,Tasinato:2023ukp,Cheng:2023ikq,Maity:2023qzw,Firouzjahi:2023bkt,Davies:2023hhn,Iacconi:2023ggt,Saburov:2024und}.

Finally, we mention possible future directions.
1) Our approach is based on the expressions in spatially-flat gauge, while the approaches in most of the previous works are based on those in comoving gauge.
We leave the discussion on how these approaches are related and on the origin of the difference in the conclusions for future work.
2) Given that we have shown the coincidence of the curvature perturbations during the first and the second SR periods, it would be worthwhile to discuss whether the conservation of the curvature perturbations is valid even during the intermediate period.
3) Throughout this work, we have neglected the slow-roll suppressed contributions. The analysis of those contributions is left for future work. 
4) The generalization to higher-order loop contributions would be an interesting future direction.

\vspace{5pt}
\noindent
\emph{{\bf Acknowledgments.}}
The author thanks Matteo Braglia, Xingang Chen, Wayne Hu, and Jason Kristiano for helpful comments on a draft of this letter.
The author was supported by JSPS Postdoctoral Fellowships for Research Abroad.

\small
\bibliographystyle{apsrev4-1}
\bibliography{draft_superhorizon_one_loop}

\end{document}